\documentclass{aa}
\usepackage{graphicx}
\newcommand{\bhm}{}
\newcommand{\ehm}{}
\newcommand{\beq}{\begin{equation}}
\newcommand{\eeq}{\end{equation}}
\newcommand{\mr}{\mathrm}
\begin{document}

\title{Striation and convection in penumbral filaments}

\author{ H. C.\ Spruit$^1$, G.\ B.\ Scharmer$^{2,3}$,  and M.\ G.\  L\"ofdahl$^{2,3}$}
\authorrunning{Scharmer,   L\"ofdahl \& Spruit}
\offprints{henk@mpa-garching.mpg.de}
\institute{Max Planck Institute for Astrophysics, Karl-Schwarzschildstr. 1, D-85748 Garching \and Institute for Solar Physics,  Royal Swedish Academy of Sciences, AlbaNova University Center, 10691 Stockholm \and Stockholm Observatory, Dept. of Astronomy, Stockholm University, AlbaNova University Center, 10691 Stockholm }
\date{Received:Accepted}

\abstract{Observations with the 1-m Swedish Solar Telescope of the flows seen in penumbral filaments  are presented. Time sequences of bright filaments show overturning motions strikingly similar to those  seen along the walls of small isolated structures in the active regions. The filaments show outward propagating striations with inclination angles suggesting that they are aligned with the local magnetic field.  We interpret it as the equivalent of the striations seen in the walls of small isolated magnetic structures. Their origin is then a corrugation of the boundary between an overturning convective flow inside the filament and the magnetic field wrapping around it. The outward propagation is a combination of a pattern motion due to the downflow observed along the sides of bright filaments, and the Evershed flow.  The observed short wavelength of the striation argues against the existence of a dynamically significant horizontal field inside the bright filaments. Its intensity contrast is explained by the same physical effect that causes the dark cores of filaments, light bridges and `canals'. In this way striation represents an important clue to the physics of penumbral structure and its relation with other magnetic structures on the solar surface.  We put this in perspective with results from the recent 3-D radiative hydrodynamic simulations. 

}

\maketitle

\section{Introduction}

The structure of sunspot penumbrae has posed long-standing puzzles. In early observational work (e.g.\ Mamadazimov 1977) the mix of long dark and bright filaments has been interpreted as showing a magnetic field (dark, as in the umbra) on top of the normal photosphere (shining through in the bright filaments). This explained the general appearance of the penumbra and the nearly photospheric brightness of the bright structures, but required that penumbral magnetic field only touches over the photosphere,  so that its optical depth would only be of order unity ( a `thin penumbra'). The field would have to be essentially horizontal in  such a  penumbra, with field lines crossing the solar surface only in the umbra. This does not agree with the observed inclination of the penumbral field (with angles varying from 10 to 40 degrees to the horizontal). In fact, most of the magnetic flux of a spot crosses the surface through the penumbra, not the umbra. The penumbral field, on average, is not horizontal, and  since $\mr{ div}\,{\bf B}=0$, the observed field must then continue to some depth below the surface. 

This has led to interpretations in terms of convection in a magnetic field extending to a substantial depth below the surface (a `thick penumbra'). An influential conceptual picture was Danielson's (1961) model of convective `rolls': an overturning flow in a plane perpendicular to a horizontal magnetic field.   Extensions of this idea to fields inclined at a finite angle to the horizontal  have led to a `magnetoconvection' view of the penumbra, which interprets the observed structures as turbulent fluctuations in a mean magnetic field  extending to depths of several 1000 km or more  (see e.g. references in Tildesley \& Weiss 2004). 

\subsection{The heat flux problem}
An important constraint on penumbral models is the well known `heat flux problem'. The bolometric brightness of the penumbra corresponds to a heat flux some 75\% of the normal photospheric heat flux. This heat is carried to the surface by convective flows, and vertical velocity amplitudes close to those observed in photospheric granulation are needed to supply this 75\%. The velocities actually observed in penumbrae are a factor of a few smaller\footnote{Typical velocity amplitudes in the granulation are of the order 1 km/s. In the penumbra, vertical velocities are harder to measure because of crosstalk from the large horizontal (Evershed) velocities; values of 100-200 m/s are quoted (e.g.\ Rimmele 1995), a factor 5 lower than in granulation. A better measure of the heat flux carried by the flows is the intensity-weighted vertical velocity. In the granulation this is $\sim 400$ m/s (e.g.\ Pierce and Beckeridge 1974). The literature does not record a corresponding measurement for the penumbra. If the degree of correlation between intensity and vertical velocity is assumed as in granulation,  the intensity weighted upflow velocity would be 50-100 m/s in the penumbra, corresponding to a heat flux of 15-25\% of the normal solar surface flux. This is a factor 3-5 lower than the measured heat flux from the penumbra.} 

 The velocity information derives mostly from parts of the contribution function above  the level where most of the energy flux is radiated.  The velocities inferred from the spectral lines thus  differ from those carrying the heat to the surface. This difference is not important in the quiet Sun because the granulation velocity amplitude drops rather slowly with height  because of convective overshoot, so the velocities deduced from spectral lines are still representative of the energy carrying velocities at the photospheric level. In the atmosphere of the penumbra, however,  the situation is different.  At the low densities where spectral lines are formed the magnetic field strongly interferes with vertical flows.  The low Doppler signals can thus be understood as a simple consequence of the strong magnetic field of the penumbra suppressing convective overshoot in the atmosphere. 

The observed penumbral heat flux, however, still requires the presence of strong upflows (velocities of the order of quiet Sun granulation) to supply the heat flux radiated by the atmosphere. A thin penumbra does this automatically, since it assumes conditions similar to quiet Sun granulation directly below the observed surface: it does not have a heat flux problem. A thick penumbra however, with subsurface field strength of the same order as the observed surface fields, has a problem because such a strong field interferes with the convective heat flux: not only in the atmosphere, but also in the immediate subsurface layers where the radiative heat flux declines steeply with depth because of the high opacity of the partially ionized gas  (c.f. discussion in Spruit and Scharmer 2006, hereafter Paper I) .

The penumbra thus show a curious mix of observational indications, somehow pointing simultaneously to two incompatible interpretations (the `thick' and the `thin' penumbra). This invites a closer look at these observations.

\subsection{Observational connections}
Sunspot observations are naturally described in terms of a classification of different structures. On closer inspection some of these phenomena turn out to be closely related to each other and also to flux  concentrations outside  spots, such as the magnetic structures in faculae (Scharmer 2009). 

This has been noted occasionally, but the fact that it provides important clues about their origin has not been exploited much before these connections were pointed out in Spruit and Scharmer (Paper I), Scharmer and Spruit (2006, hereafter Paper II). 

The heads of bright filaments extending into the umbra regularly turn into umbral  dots  (e.g. p137 of the review by Zwaan 1968). The `dark cores' over bright  filaments (Scharmer et al. 2002) are also very prominent in light bridges. In fact, light bridges can form continuous connections with penumbral bright filaments (e.g.\  Beckers and Schr\"oter 1969, Langhans, 2006 (see the movies `Light bridges' and 
`Sunspot' in the online-only Appendix\footnote{also available at the site of the Institute for Solar Physics (ISP):\hfill\- {\tt http://www.solarphysics.kva.se/gallery/movies/ oslo-2004/movies/gband\_20Aug2004\_sunspot\_41min.mpg}\hfill\break and {\tt http://www.solarphysics.kva.se/gallery/movies/ dark-cores-2002/full\_color.mov}}), Rimmele 2008, Katsukawa et al. 2007), and the two can evolve into each other. 

The interpretation of light bridges as inclusions of \bhm photospheric convection \ehm  embedded in an umbra is  well documented and appears uncontested in the literature.  For example, the gradual evolution of light bridges into normal photospheric convection as part of the decay process of spots was described very early in the history of the subject (e.g.\ Bray and Loughhead 1964, Vasquez 1973 and references therein). Their formation during the growth of spots, as remnants of inclusions of normal photospheric surface, has also been described in detail in e.g.\  Vrabec (1974), Bumba and Suda (1983).    

A well-developed explanation of umbral dots is that of Parker (1979). In this idea, dots are caused by gaps in the umbral field which open just below the observed surface of the umbra.  Observations supporting the interpretation of light bridge phenomenology in terms of this idea have been presented among others by Kusoffski and Lundstedt (1986), Garc{\'\i}a de la Rosa (1987), Sankarasubramanian and Rimmele (2002), Jur{\v c}{\'a}k et al. (2006). It agrees also with the well-known observation that light bridges often decay into strings of umbral dots (e.g.\  Zwaan 1968, Louis et al. 2008). 

Apart from this observed connection, the interpretation of dots in terms of sub-surface gaps  provides the only realistic solution to the umbral heat flow problem. As in the penumbra, the heat flux in the umbra, about 20\% of the normal solar surface flux, can not be carried by the observed vertical velocity amplitudes (Beckers 1977). In Parker's umbral gap model, the umbral heat flux is carried by field free convection in the gaps, which close near the continuum optical depth unity surface or below.   In contrast with normal granulation, the flow is therefore absent above the continuum level, in the region where the magnetic field dominates and suppresses convection (cf. Sankarasubramanian and Rimmele 2002). This explains the low velocity amplitudes often observed in dots, both in net velocity and in line broadening (Beckers 1977, Adam 1979, Schmidt \& Balthasar  1994, Hartkorn \& Rimmele 2003). As expected in the gap picture, however, this depends on the brightness of the dots: faint ones (i.e. the ones closing deeper down) show little or no velocity signal, while bright ones show velocities similar to those in light bridges (from which they often evolve). 

This interpretation has been confirmed with realistic 3-D radiative MHD simulations of a sunspot umbra by Sch\"ussler and V\"ogler (2006). These show the formation of umbral dots, including the dark cores crossing them that have been observed in large dots (for recent references see Bharti et al. 2007; Rimmele 2008; Riethm\"uller et al. 2008).

These observational connections form the basis for our interpretation of penumbral structure as also consisting of gaps: opening below the surface just like umbral dots, but elongated along the horizontal component of the penumbral field. This interpretation explains the observed heat flux, the origin of the dark cores over bright filaments, and the variation of field strengths and inclination in penumbral structure (Paper I, Paper II, Scharmer 2008, for a review see Scharmer 2009).

\begin{figure}[h]
\centerline{\includegraphics[width=0.8\hsize]{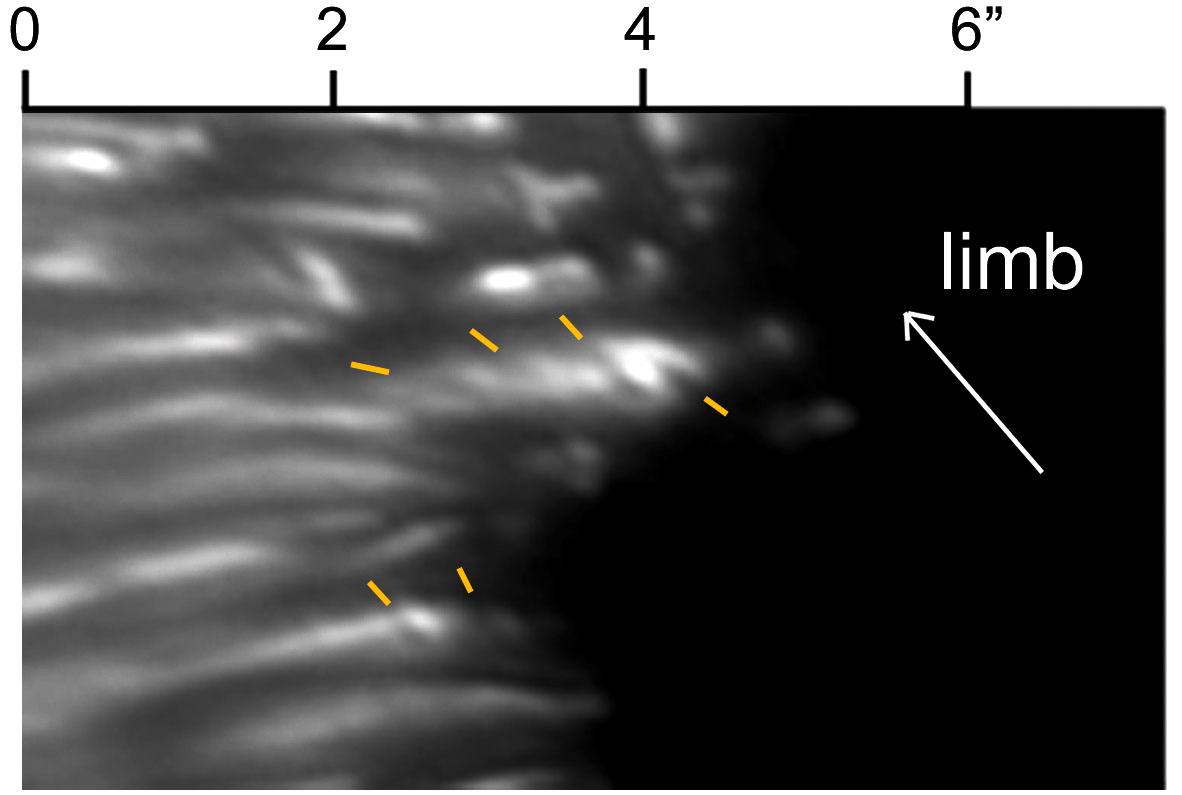}}
\caption{
Part of a large spot observed with 1-m Swedish Solar Telescope on 2 May 2003. The heliocentric distance  from disk center is 70$\degr$  ($\mu=0.35$), arrow indicates direction to the limb). Striations on two filaments are marked. Overturning motions are seen most clearly in the large filament at the center, see the video clip `Striation' in the online material (also available at {\tt http://www.mpa-garching.mpg.de/$\sim$henk/stria.mp4}). The striations are propagating to the left (away from the umbra).}\label{stria}
\end{figure}

The interpretation sketched above thus predicts that the flow in a bright filament should have the characteristic pattern of overturning motions seen in granulation. The observations presented below confirm this prediction. 

On top of this overturning flow exists the conspicuous outward Evershed flow. This is a topic in itself we will not address here, except to note that it has also been observed already in realistic 3D radiative MHD simulations. These numerical results show that it can be understood as the horizontal flow component of this form of gappy convection (Scharmer et al. 2008).

The striation pattern was already clear from observations made with the SST in  2003 (unpublished). In the following we analyze these observations, which show the overturning motions in great detail at a resolution approaching 0\farcs1. We compare them with overturning flows seen in the granulation around small isolated magnetic structures (pores). In both cases, substructure in the form of striation is seen. The observational properties of this striation are discussed.   Its structure is interpreted as a corrugation (fluting) of the boundary between the penumbral magnetic field surrounding the gap and the flow inside it, and  compared with the results from recent 3-D radiative MHD simulations.

Observations with the Hinode telescope reported by Ichimoto et al. (2007) also show the striation pattern. These authors did not give a clear interpretation,  but  nevertheless suggested that it is due to `twisting motions' of the magnetic field, a view that also dominates later references.  A more explicit interpretation was made by Zakharov et al. (2008) using observations with the Swedish 1-m Solar Telescope (SST).  They show that a  twisting field explanation conflicts directly with the observed motion of the pattern, which is systematically away from the solar limb. Whichever way a magnetic field might be twisting, it does not know about the solar limb seen by the observer. They interpret the pattern sensibly in terms of a convective flow. As we shall argue below, a  convection pattern is indeed indicated by the observations, but not in the form of the magnetic `rolls' proposed by these authors. Instead, the flow must be in the form of an at best weakly magnetic, overturning form of convection more akin to that seen in the granulation.

\section{Observations}

 \begin{figure*}[t]
 \includegraphics[width=0.9\hsize]{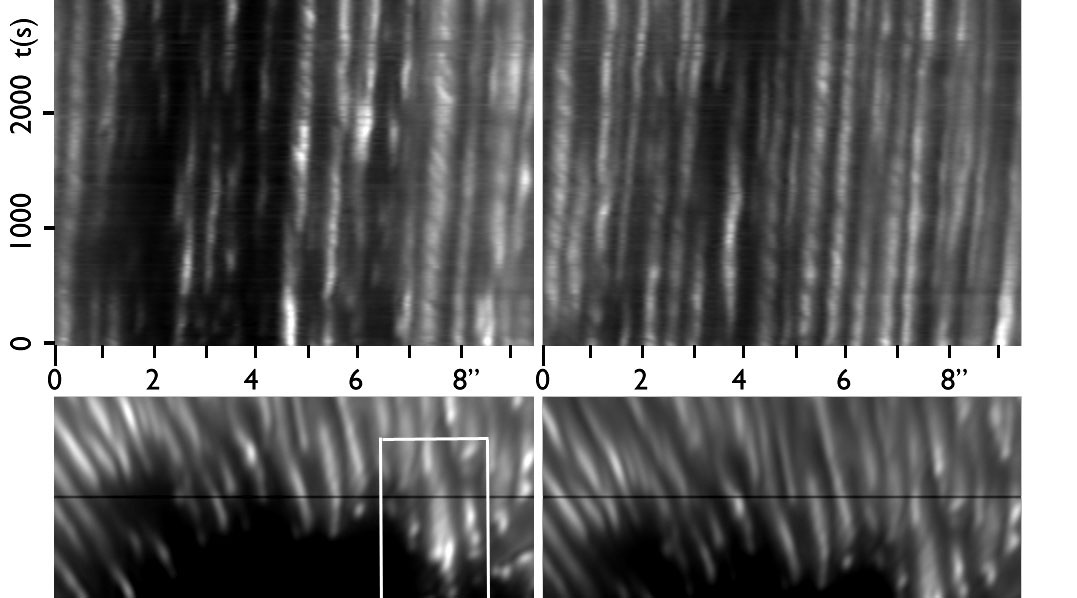}
 \caption{Intensity as a function of time along a slit perpendicular to the filaments (at two positions shown in lower panel).  The slanted structure  at about   45$\degr$ crossing the bright lanes corresponds to motion of the striation at a speed of  2  km/s perpendicular to the filaments. (The slight inclination of the lanes themselves reflects the image drift due to the Sun's rotation).  White rectangle indicates approximate location of the filaments in fig. \ref{stria} (rotated 90$\degr$ clockwise, and at a different time in the sequence).} 
 \label{perpslit}
 \end{figure*}

Fig.\ \ref{stria} shows a part of the penumbra of a large symmetric spot. The field is from a time series of images taken on 2 May 2003 of the spot in AR351 (heliographic coordinates N8 E65, $\cos\theta\approx 0.53$).  Image data were collected through a 430.5 nm G-band interference filter with a MegaPlus 1.6 1534 x 1024-pixel camera at 13 ms exposure. The camera was equipped with a phase diversity beamsplitter, so that half the detector recorded a conventional in-focus image, while the other half recorded a simultaneous but intentionally defocused image of the same field of view. The SST Adaptive Optics system was running and frame selection was used, storing the best 3 frames in 20-sec intervals. The residual seeing effects were further reduced by use of Joint Phase Diverse Speckle (L\"ofdahl 2002) on the selected images, resulting in a sequence of restored images with an average cadence of 20 sec. These images were corrected for image rotation and anisoplanatic warping.

\subsection{Striation}

Well-defined bright filaments  show a clear striation: dark lanes at angles of 10-45$\degr$ to the filament axis. Bright filaments are most clearly defined as individual structures in the inner part of the penumbra, and the striation is also seen most clearly there. The field shown in Fig.\ \ref{stria} is centered on a wide,  well resolved bright filament. The time-dependence of this striation is illustrated as a `time slice' in  Fig.\  \ref{perpslit}, showing the intensity as a function of time along a slit taken perpendicular to the filaments.  The substructure seen crossing the filaments in this image corresponds to a proper motion of about 2 km/s. 

\begin{figure}[b]
\includegraphics[width=1.03\hsize]{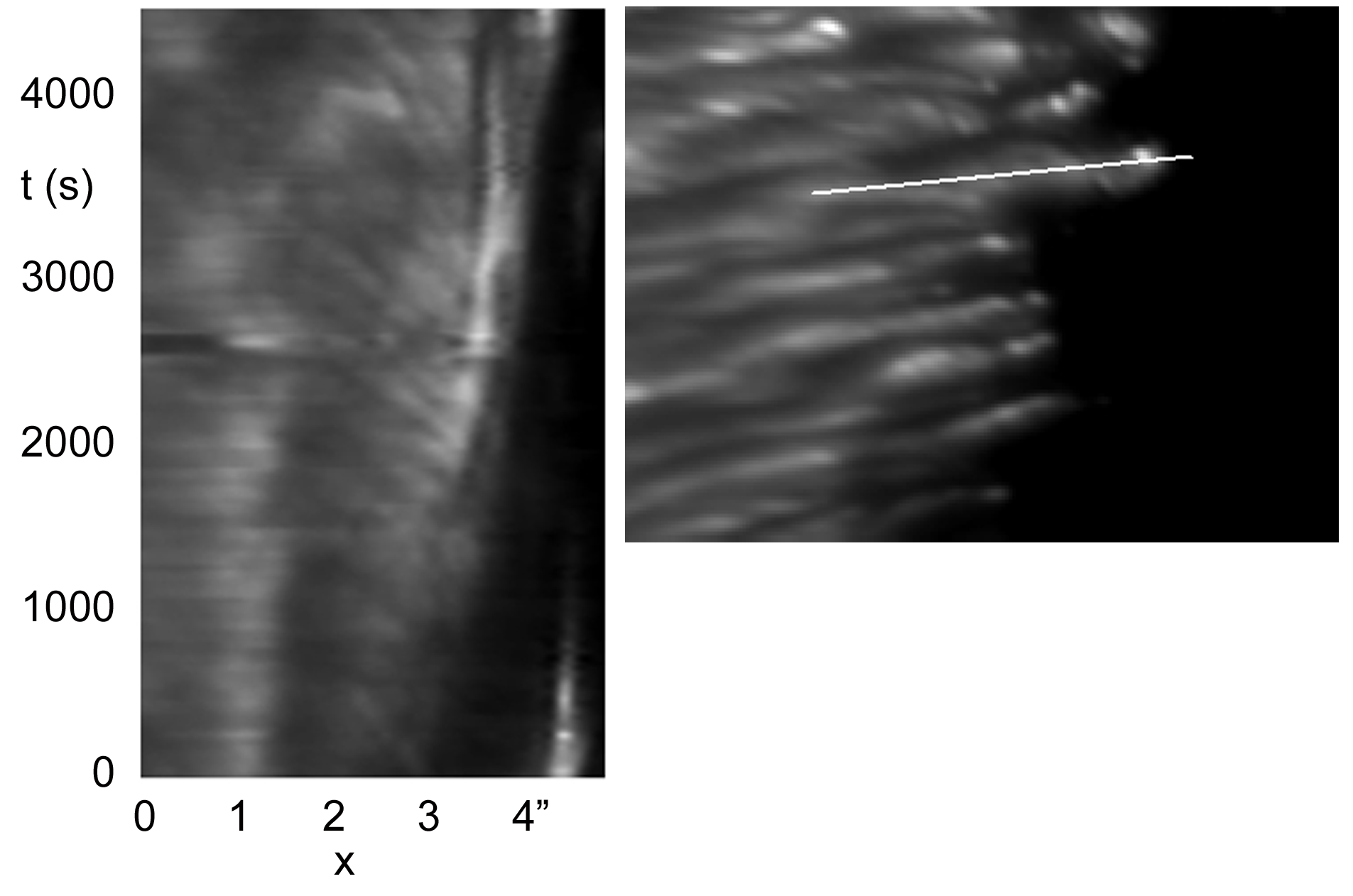}
\caption{Intensity as a function of time (left panel) along a slit (right panel) parallel to a large filament.  The slanted structures at $\sim45^\circ$ correspond to outward propagation of the striation along the filament, at speeds of 0.7 - 3 km/s. The inward drift of the filament head, slowing down with time, is seen in the right half of the left panel.} 
\label{parrslice}
\end{figure}

The motion of the striation can also be seen by taking a slit along a single filament. This is illustrated by the time slice shown in Fig.\ \ref{parrslice}. On the right (umbral) side of the image, it shows the well-know inward penetration of filament heads. The striation, seen most clearly from $x=1.5$ to $x=3$\arcsec,  shows an outward motion (away from the umbra). This is also striking as a visual impression in a movie of the time series. The apparent speed  ranges from 0.7 to 3 km/s. 

The pattern seen in Fig.\  \ref{perpslit} is also seen in the data from the Hinode satellite in Ichimoto et al. (2007). The systematic motion away from the limb  at all azimuthal angles and disk center positions excludes the interpretation in terms of `twisting motions' by Ichimoto et al. (2007) and by later authors. As pointed out in Zakharov et al. (2008)  the systematic flow away from the limb, instead, shows it to be a convective flow. These authors discuss it in terms of a `convective roll'. As we shall argue presently (see also Paper I,II), a simpler and more convincing interpretation is in terms of an \emph{ overturning} convective flow rather than a roll: much like the flow seen in ordinary granulation.  The difference is in vertical structure and energy budget: in a closed roll the thermal energy content is finite, and would suffice for the observed heat flux of a bright filament for only a few minutes before having to be replaced by another roll. The observed  life time of bright filaments argues against this (see also the discussion in Scharmer 2008, 2009). Instead, an overturning flow, deriving from and returning to large depths (as sketched in Fig.\ \ref{sketch}) can supply the radiated heat for as long as the filament exists.

\section{Overturning}

 Viewing the structure in Fig.\ \ref{stria} as a time series already gives the impression of an overturning motion, reminiscent of the downflow pattern seen around the small magnetic structures near the limb in Fig.\ \ref{oslo} discussed below. 

The striation  seen in Fig.\ \ref{perpslit} gives additional clues. Its proper motion  of about 2 km/s is a projected speed: if the surface on which it takes place is horizontal, the actual speed would be a factor $1/\cos\theta\approx 2$ higher, for the disk position of our observations. 

From their aspect at different viewing angles,  the optical depth unity surface of the bright filaments is known to be elevated above that of their surroundings (Paper I, Paper II, Ichimoto et al. 2007, Zakharov et al. 2008), as is seen also in light bridges (Lites et al. 2004).   In spots near the limb, the observer therefore sees the sides of elevated structures facing disk center (compare Fig. \ref{sketch}).  Since the elevation of bright filaments is similar to their width,  the actual viewing angle  for our observations  is probably closer to perpendicular to the optical depth unity surface, in filaments oriented roughly parallel to the limb like those in Fig.\ \ref{perpslit}  as well as those in Ichimoto et al.  The apparent flow speed is  then representative of the actual flow speed along the filament surface.

Inclined structures carried down by an overturning flow would appear to move outward (`barber pole' or `scissors' effect). For the observed inclinations, convective flow speeds of $\sim 2$km/s would produce approximately  the observed speeds. 

This is also consistent with the fact that the highest apparent speeds are seen in striation with the smallest angle with respect to the filament axis  (on average, with considerable scatter).  However, even though our observations refer to filaments in the inner penumbra, the  Evershed flow may also contribute to the outward motion. It reaches its largest speed in the outer penumbra, but speeds of $\sim 2$ km/s are found already in the inner penumbra.

\section{Interpretation of the striation}

\subsection{Connection with flows around small magnetic structures}

\begin{figure}[b]
\includegraphics[width=\hsize]{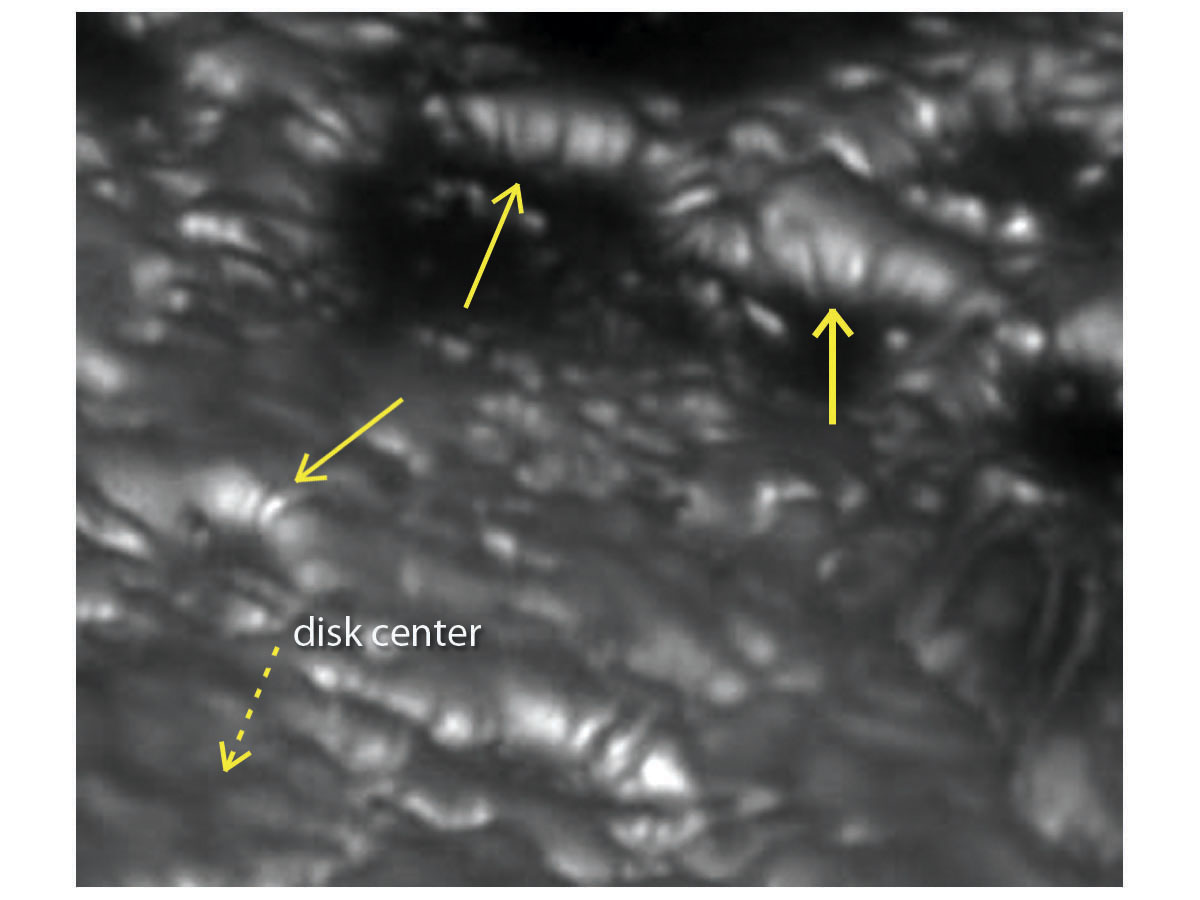}
\caption{Pores in an active region near the limb of the Sun observed with the 1-m Swedish Solar telescope on 10 May 2004. Dashed arrow indicates direction to the limb. Overturning motions with downflows are visible in the bright edges at the limb-side of the pores. A particularly clear example is the pore on the right  (heavy arrow); for a video clip showing this see `Overturning' in the online-only Appendix (also at the original location: {\tt http://www.solarphysics.kva.se/gallery/movies/ oslo-2004/movies/gband\_10May2004\_AR\_limb.mpg}). Image and movie courtesy by Luc Rouppe van der Voort }\label{oslo}
\end{figure}

 Phenomena strikingly similar to the overturning motion in the penumbral filaments presented in the previous section are seen in high-resolution observations of convection around the small magnetic elements that make up most of a young active region. An example is shown in Fig. \ref{oslo}, taken on 10 May 2004 with the SST. 

The magnetic field in such elements reduces the gas pressure, so they are more transparent and appear as `dips', or depressions in the observed surface of the Sun (Spruit 1976, 1977). This is particularly clear in observations near the limb of the Sun. The  limb-side  rim of such a dip is seen as a brightening while the proximal boundary is obscured. Radiative cooling of gas surrounding the boundary increases its density; as a consequence the element is surrounded by convective downflows. Given sufficient spatial resolution, these flows can be observed directly in time sequences of images such as Fig. \ref{oslo}. 

The phenomenology seen in such observations has been reproduced in detail by \emph{ ab initio} radiative magnetohydrodynamic simulations (Carlsson et al. 2004, Keller et al 2004, De Pontieu et al. 2006, see also Steiner 2005). We can thus be confident of the interpretation given above: we have a good understanding of what overturning convection along a magnetic boundary in the solar photosphere looks like.

The boiling, overturning impression given by the penumbral movie of Fig. \ref{stria} is similar to the flows seen on the limb side of the pores in Fig. \ref{oslo}. Apart from the overall impression, the time scales and length scales as well as the `striated' substructure are common properties. The main difference is the orientation of the striation. In the magnetic structures in Fig.\ \ref{oslo} the striation is parallel to the downward flow; in the penumbral filament it is at an angle. 

For isolated magnetic elements we know that the striation follows magnetic field lines: it is a corrugation of the surface bounding the magnetic structure from the surrounding convection zone. [This is demonstrated by comparison with the MHD simulations, e.g.\ De Pontieu et al. 2006]. The striation in the penumbral filament, on the other hand, is inclined at angles expected for the field at such a position in the penumbra. The obvious interpretation is thus that the striation is a corrugation of the magnetic surface surrounding the filament, outlining the direction of the field lines. The downflow along the boundary, carrying the corrugation with it, causes an apparent outward motion of the  striation. 

 It is likely that  a real outward fluid motion along the gap also contributes to the motion of the striation. The numerical simulations of Heinemann et al. (2007) show an outward flow in the gaps, along the boundary with the magnetic field. Scharmer et al. (2008) discuss its origin, and  conclude that this (Evershed) flow is the horizontal component of penumbral convection.

\begin{figure}
\includegraphics[width=\hsize]{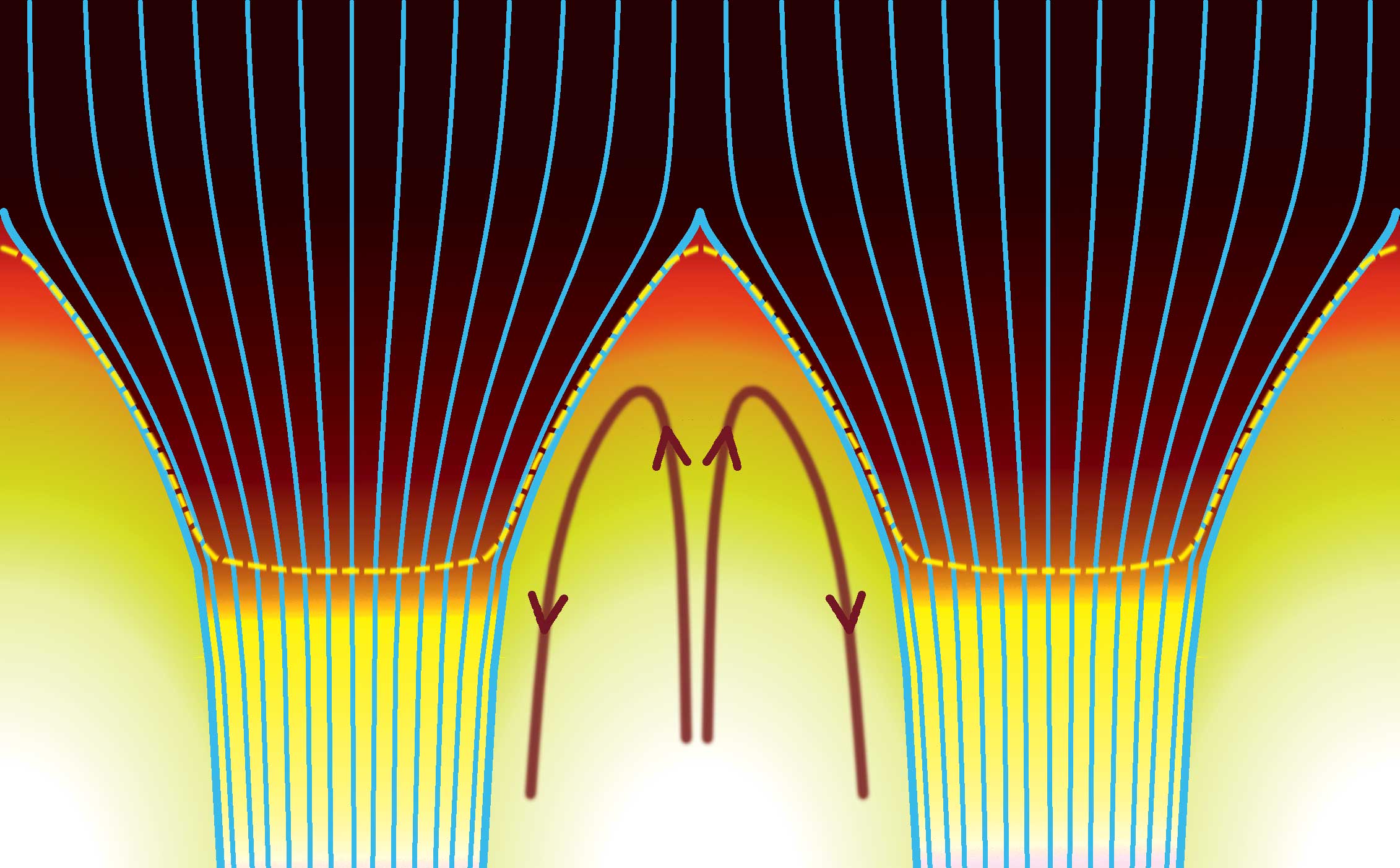}
\caption{Sketch of a convecting gap in the magnetic field of the penumbra, showing a cross-section perpendicular to the axis of the penumbral filaments. Colors in the gap indicate temperature (high temperature yellow/white, photospheric temperatures red). Gas density is indicated as brightness, increasing downward. The gap has a higher gas density than the surrounding magnetic  field. Blue: magnetic field lines projected on the plane. (The field component perpendicular to the plane is not shown). Yellow dashed line: optical depth unity surface (as seen from the top in the continuum). The cool higher density gas at  the top of the gap makes it appear as a dark core overlying the filament. Between the gaps convection is suppressed by the magnetic field, which therefore also appears dark. (Field configuration taken from the calculations in Paper II)} 
\label{sketch}
\end{figure}

\begin{figure}
\includegraphics[width=\hsize]{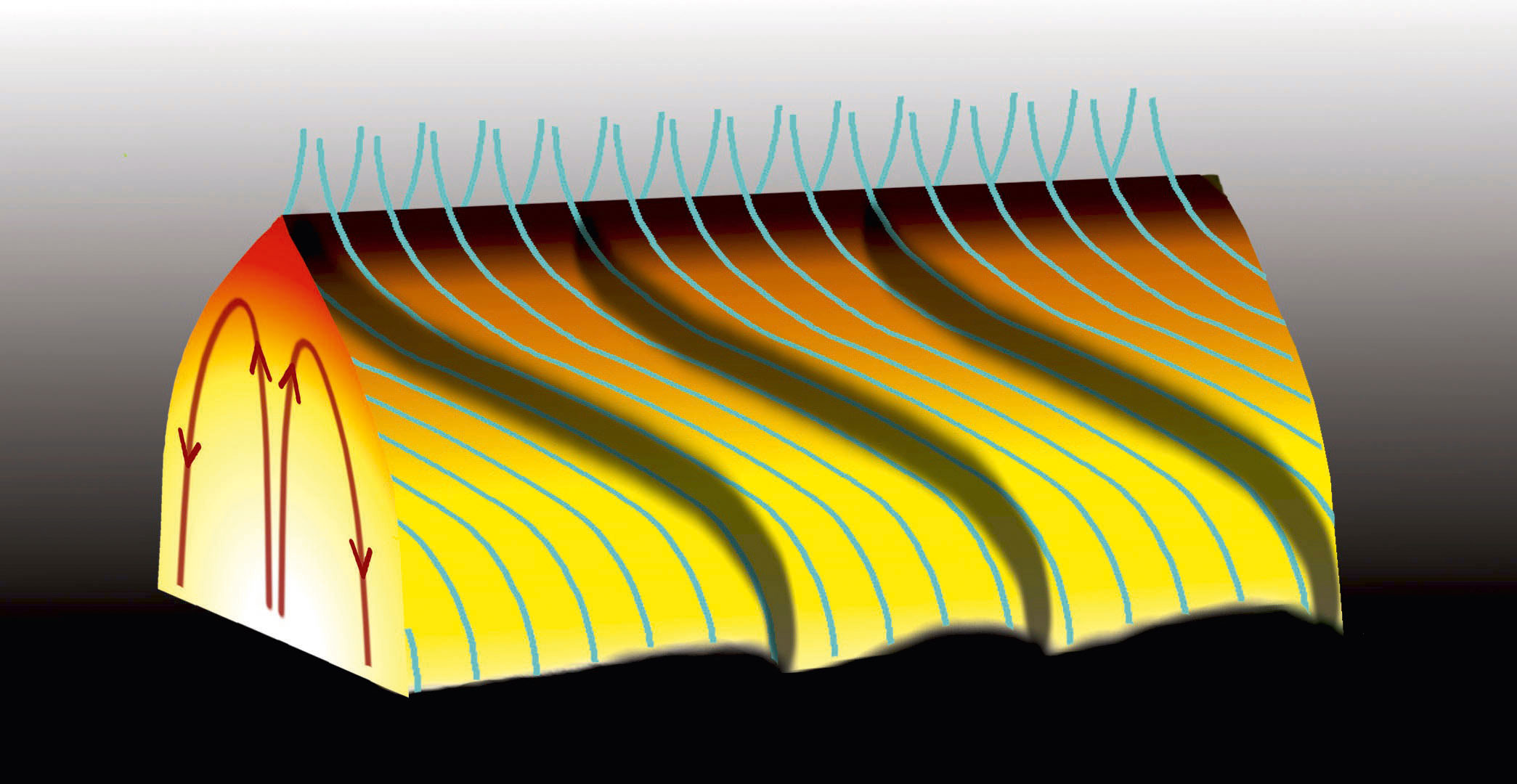}
\caption{Perspective drawing of a convecting gap in the magnetic field of the penumbra. Sketched is the surface of the gap as it would be seen in the continuum, with a dark core over the top of the gap, and dark striation of the surface parallel to the field lines (blue) wrapping around the gap. 
Direction to the umbra is to the right.  The horizontal (Evershed) component of the flow is not shown.} 
\label{3dsketch}
\end{figure}

\section{Convection in penumbral filaments}
\subsection{Convective expulsion}
Overturning convection such as seen in the movie of Fig.\ \ref{oslo} keeps magnetic flux separated from its environment by the mechanism of \emph{ convective flux expulsion} (Zel{'}dovich 1956, Parker 1963, Weiss 1966).  It is the process whereby the small scale field on the surface remains concentrated in intergranular lanes. Diffusion of the magnetic field into its environment is matched by  the convective flow advecting the field lines back into the field concentration. The balance between the two processes determines the thickness of the boundary separating the flow from the field.

The speed of the overturning flow inferred from the data in Fig.\ \ref{perpslit}, about 2 km/s, agrees well with the velocities found in the simulations of Heinemann et al. (2007).

\subsection{`Field free'}
The question can be raised when convecting flows such as seen in Figs. \ref{stria},\ref{oslo} \bhm are `field free' as in Parker's view of umbral dots\ehm. A convecting flow near a magnetic boundary will at some level carry stray fields with it,  resulting from diffusion or hydrodynamic entrainment of the external field. \bhm This would broaden the transition between the flow-dominated interior and the surrounding magnetic field, so that observed magnetic signals could look the same as those of a fully magnetized region\ehm. The physics of flux expulsion, however, makes the distinction between \emph{ convecting} and \emph{ magnetically dominated} regions conceptually unambiguous.  \bhm In regions where the field is weak enough that the kinetic energy of the flow \ehm dominates over the magnetic forces, the flow behaves like convection in the absence of a field, as in the photosphere outside locations of strong field and in the classical flux expulsion calculations by Weiss (1966).  This is the sense in which the gaps in Parker (1979) and in Papers I,II are called `field free'. In fact, the properties of the observed striation can be used to derive an upper limit to the strength of the magnetic field inside the \bhm main body \ehm of a bright filament; this is discussed in section \ref{limfield}.

\subsection{The corrugated boundary}
\label{corrug}
\subsubsection{Fluting}
\label{fluting}
Magnetic fields have cohesion only along the field: neighboring field lines can slip parallel to each other without restoring forces. The surface bounding a magnetic region from its surroundings is therefore easily corrugated or `fluted' (like the columns of Greek temples). 

In fact, a magnetic boundary will corrugate spontaneously by a \emph{ fluting instability} if the curvature  vector  of the field lines points into the external medium (e.g.\ Bateman 1980; for an analysis in the context of sunspots see Meyer, Schmidt and Weiss 1977). This is the case with the boundaries of the pores seen in  Fig.\ \ref{oslo}, because the field fans out above the surface.  Fluting is thus a  likely  explanation  for the striation seen in the walls of small magnetic structures. 

Field lines wrapping around a gap as in Fig.\ \ref{sketch} also have a fluting-unstable curvature.  As suggested in Paper II (section 7) the ubiquitous striation seen in penumbral filaments is thus plausibly  due to fluting.  If conditions for fluting instability are only marginally satisfied, external forcing by `noise' in the form of irregularities in the flow inside the gap would contribute as well. 

 The presence of the striation is an argument against the existence of a longitudinal (filament-aligned) magnetic field in the gap (such as proposed by Zakharov et al. 2008). The displacements of the gap boundary due to fluting would bend such a field. The restoring tension forces resulting from this bending oppose fluting. This is exploited in the design of controlled fusion devices by  judiciously shearing the field line directions across the magnetic surfaces, which would otherwise be unstable to fluting. This stabilizing effect is strongest at the shortest wavelengths. The very narrow widths observed in the striation (at the resolution limit) thus argue against a longitudinal field of significant strength in the gap.

\subsubsection{Source of the brightness contrast}
How does such a corrugated surface cause  the brightness contrast observed as striation? The corrugation develops slowly compared with the relevant sound (fast mode) crossing times, hence it  must take place approximately in \emph{ pressure equilibrium}: $P+B^2/8\pi=$ cst., both in time and across the boundary.  Gap fluid, with low field strength and high plasma density, forms ridges protruding into the magnetic field where the density and optical depth is low. Radiative losses from such a ridge are larger than in its surroundings. The resulting steepening of the temperature gradient causes the ridge to appear dark. 

The physical conditions giving rise to the contrast seen as striation of penumbral filaments are thus the same as in the dark cores over penumbral filaments, in light bridges  (see Nordlund and Scharmer 2009)  and in the dark striation in small magnetic elements seen in Fig.\ \ref{oslo}.  Since the field lines wrapping around the gap continue into the dark core over a filament, this interpretation of striation contrast  also accounts for the observed   close connection of  striation with the dark cores overlying filaments. It is also consistent with the analysis by Carlsson et al. (2004) of the striation seen in their simulations of small magnetic structures.

\subsection{Optical depth to the boundary}
\label{depth}
For a field strength of 1500 G (mid-penumbra) the magnetic pressure $B^2/8\pi\approx 10^5$ erg/cm$^3$. This matches the photospheric pressure fairly closely. The boundary between field and gap will thus occur close to a (continuum) optical depth unity. It is conceivable that most of the line formation takes place in the magnetic volume, with little contribution from the \bhm gap interior\ehm. In this case it would be impossible to detect the magnetic boundary through its effect on the polarized line profile. 

On the  other hand, if the density above the boundary is low enough, the optical depth to the boundary is low, so the line profile will be formed in part \emph{ below} the boundary, resulting in a weaker polarization signal. Observationally, this would have an effect similar to a `stray light' contribution,  and a lower apparent field strength. 

The elevation of bright filaments over their surroundings, of the order 300 km (as seen in the numerical simulations,  and consistent with the changing aspect of observed filaments with viewing angle) is about two pressure scale heights at the observed temperature of the filaments. The time scales for changes in the filaments (10-30 minutes) are longer than the sound travel time over such a height (about 40s). The gas pressure on the field lines bounding the gap must therefore decrease with height approximately according to  hydrostatic equilibrium.  Near the top of the gap the pressure on these field lines would then be a factor 10 lower than at the optical depth unity level between the gaps (cf. Fig.\ \ref{sketch}), and the optical depth to the magnetic boundary would be correspondingly lower.  One would expect this difference to be reflected in the polarization signals: it would reduce the inferred field strength in the dark cores (already low because of the cusped nature of the field configuration).

\subsubsection{Thickness of the boundary between field and gap}

 The magnetic signals to be expected depend critically on the thickness of the boundary between the gap and the field surrounding it. If only Ohmic diffusion plays a role, the transition may be as thin as a few km, which would not have much consequence for  spectral line formation. The depth into the gap where optical depths $>2$ are reached is only $\sim 20$ km, however, so only a mild increase of the penetration of the magnetic field into the gap could have significant effects on the expected polarization signals.  For comparison, this is of the order of the grid resolution in current numerical simulations of sunspots. Simulations sufficient for realistic line formation calculations will need significantly higher resolution.

A complication in this context is the possibility that the transition between the magnetic field and the convecting flow in the gap may be broadened by the consequences of fluting. The smallest length scale perpendicular to the field lines is limited only by microscopic diffusion, and the growth rate of the instability is independent of the length scale as long as it exceeds this limit.   If  the available time (the flow time along the boundary) is sufficient  for the instability to grow significantly, small length scales are thus likely to develop. If this happens the transition may be significantly broadened by the effect of Ohmic diffusion on these small scales.  External forcing such as Kelvin-Helmholtz instability will have the same effect.

Striations on length scales below the observational resolution are thus likely to be present if the observed striation is indeed caused by fluting instability. In addition to broadening the transition, corrugation on such short length scales would provide an efficient lateral path for radiative exchange into the magnetic region, adding another scattered light component to the expected magnetic signals.

\subsubsection{Limit on field strength inside the gap}
\label{limfield}
 The striation indicates field line directions at substantial angles to the filament. A strong field inside the filament,  as in the `rolls' picture of Zakharov et al. (2008) does not agree with this observation. Substructure in such a roll would instead be parallel to the filament. 

If the striation is due to the field surrounding the filament, as we propose here, it can be used to put a limit on field strength inside the filament. Corrugation of the boundary seen as the striation would bend the field lines inside, since they are at a different angle. The restoring force due to this bending opposes the corrugation. This can be made quantitative as follows.  Let the width of the filament be $d$, the smallest wavelength of the striation $\lambda_\mr{ min}$, the field strength surrounding the filament $B_\mr{ e}$, the field inside it $B_\mr{ i}$. The condition that the force driving corrugation can overcome the restoring force of the internal field is equivalent  to the condition that its growth rate $\sigma$ exceeds the frequency of an Alfv\'en wave of wavelength $\lambda_\mr{ min}$ in the internal field. The growth rate is $\sigma\approx B_\mr{ e}/(4\pi\rho_\mr{ i})^{1/2}/r_\mr{ c}$ where $r_\mr{ c}\approx d/2$ is the radius of curvature of the external field ${\bf B}_\mr{ e}$, and $\rho_\mr{ i}$ the gas density inside the filament (see Meyer et al. 1977. Gas density inside ${\bf B}_\mr{ e}$ has been neglected). This yields
\beq 
{B_\mr{ e}\over (4\pi\rho_\mr{ i})^{1/2}d/2}>{2\pi\over\lambda_\mr{ min}} {B_\mr{ i}\over (4\pi\rho_\mr{ i} )^{1/2}},
\eeq
i.e.\beq B_\mr{ i}< {\lambda_\mr{ min}\over\pi  d}B_\mr{ e}. \eeq
The shortest wavelengths seen in the striation (limited by the telescope resolution) are about half the width of the filaments, so the above argument gives an approximate upper limit of $B_\mr{ i}/B_\mr{ e}\approx 0.2$. For a field strength $B_\mr{ e}\approx 1500$ G characteristic of the inner penumbra, this yields $B_\mr{ i}\la 300$ G.  Such a field is dynamically weak, and its contribution to observed polarization signals would be small.

\section{Summary}
 
A striation pattern is observed in penumbral filaments, especially in filaments oriented along the solar limb in spots seen away from disk center. 

In the direction perpendicular to the filament, it propagates towards disk center, and outward (away from the umbra) in the parallel direction. Rather than `twisting' of the magnetic field, this pattern reflects an overturning convective flow inside the filaments. The overturning motion is also observable directly in the best resolved filament in our data; it looks very similar to the overturning seen in the photosphere around small scale magnetic structures like pores seen toward the limb. 

The striation itself can be understood as a natural consequence of the fluting tendency of magnetic field lines wrapping around a region of low field strength (`gap'), causing a corrugation of the interface. The brightness contrast of the striation can be understood as a consequence of excess radiative loss in the parts of the corrugated surface protruding into the magnetic field. 

Since a magnetic field inside the filament  would stabilize its boundary against such fluting, the presence of striation sets an upper limit on the strength of such an internal magnetic field.   

Fluting may broaden the  interface between the field and the gap; the exact width of the transition is likely to be important for the observed polarization signals.

\section{Discussion}

The observations presented provide further support for the picture of bright penumbral filaments as elongated gaps in the magnetic field,  containing the same kind of overturning convecting flows observed elsewhere on the solar surface. This picture does not connect with the classical magnetoconvection view of penumbral structure as turbulent fluctuations in a mean field. 

It connects very well, however, with the known physics of magnetic flux expulsion by flows in highly conducting fluids like the solar convection zone. The clear separation between magnetic and nonmagnetic areas observed in the small scale fields at the solar surface is well understood in terms of this process. There is no good reason why this separation should not  operate in sunspots as well.  

Light bridges and umbral dots are accepted as phenomena that fit this picture  (e.g.\ Vazquez 1973, Parker 1979, Garc\'{\i}a de la Rosa 1987, Leka 1997), and  the step from there to penumbral structure is actually not a very dramatic one. The main obstacle to this step is that it conflicts with a long standing tradition of interpreting sunspot structure (for a review of this tradition see Thomas and Weiss  2004). 

The basic correctness of this `gappy' picture has been confirmed in remarkable detail already by realistic numerical simulations of small spots by Heinemann et al. (2007) and Rempel et al. (2009). These reproduce the overturning flow speeds seen in data like Fig.\ \ref{perpslit}, the inward motion of penumbral filaments, the dark core phenomenon discovered by Scharmer et al. (2002), the varying aspect of penumbral structure with viewing angle, the moat flow, and the Evershed flow (Scharmer et al. 2008).  

The realism required to achieve such comparison with observations is easily  missed by ignoring any of several pieces of physics that,  though not important in the opaque gas pressure dominated deeper layers, become crucial at the observed surface.  To reproduce the structure of the magnetic field at the surface of a spot, a proper treatment of the magnetic field in the tenuous, low-$\beta$ atmosphere above the surface is critical. The high Alfv\'en speeds here strongly restrict the kind of magnetic configurations that are possible near the observed surface (cf. discussion in Paper I).  Most of the traditional `magnetoconvection' experiments miss this point by leaving out the magnetically dominated atmosphere altogether.  Attempts to interpret sunspot structure by analogy with such models, or interpretations based on magnetic turbulence formalisms can not be expected to add much to understanding of observed sunspot structure.

The response of the atmospheric field  is fast compared with the changes taking place at its photospheric boundary. As a result it takes place  approximately along a series of  minimum energy states corresponding to the changing boundary conditions. The peculiar pattern of variations in field strength and inclination, which observers have interpreted in terms of thin floating flux tubes, is simply  the expected response  of the atmospheric magnetic field to the opening of a gap between the field lines below the surface,  aided by surface cooling of horizontal convective (Evershed) flows along the filaments (see discussion in Nordlund and Scharmer 2009).

Flux tubes suspended in such a magnetically dominated atmosphere, while computationally convenient as a one-dimensional reduction, are physically unrealistic non-equilibrium structures. It is not surprising that nothing like tubes (twisted or otherwise) turns up in the numerical simulations. At the same time, the observations leave less and less room for these constructions, as the spatial resolution achieved with improving technology increases (Scharmer 2009). 

 Much effort has been devoted to inversion of spectropolarimetric observations into (magnetic) atmospheric structure models. Such inversions are notoriously poorly constrained. They are regularized in practice by imposing an assumed structure on the field configuration, such as the popular embedded flux tubes proposed first in the `uncombed' model of Solanki and Montavon (1993). Such inversion produces answers whether or not there is a sound physical basis for the assumed structure, however (for example, assumptions violating $\mr{ div}\,{\bf B}=0$, c.f.\ Borrero et al. 2006).   Fits obtained in this way thus give a misleading sense of confirmation of the input models.

Zakharov et al. (2008) propose to accomodate classical Danielson rolls within a gap model by placing them inside the gaps. This provides a sense of continuity with traditional views of the penumbra. It also retains the flux tubes proposed earlier, but moves them from their physically awkward position in the atmosphere to a place below the observed surface. The gaps proposed in Paper I,II already explain the observations well without such additions, however,  and the addition does not agree well with the magnetic expulsion process of convective flows. In addition (section \ref{fluting}),  a longitudinal field of any significant strength in the gaps would suppress any corrugation of the filaments, especially on the very short wavelengths actually seen in the striation.

Next to the treatment of the atmospheric magnetic field,  the physics  of radiation is of equal importance  for realism in numerical simulations. Cooling by radiation at the surface determines the thermal structure of the penumbra and drives the observed flows. On the other hand, it also determines the detailed appearance of penumbral structure at the optical depth unity surface. Any physically meaningful comparison with observations thus requires inclusion of radiation physics at a fairly well developed level. 

The fact that the required level of realism in the treatment of magnetic fields and radiation physics has now been reached, and a significant degree of convergence with observations already achieved, can count as a major breakthrough in solar physics. 

\begin{acknowledgements}  It is a pleasure to thank Dr.\ Schlichenmaier for his critical comments, which have led to extensive improvements in the presentation. We thank Luc Rouppe van der Voort for allowing the use of Figure 4. The Swedish 1-m Solar Telescope is operated on the island of La Palma by the Institute for Solar Physics of the Royal Swedish Academy of Sciences in the Spanish Observatorio del Roque de los Muchachos of the Instituto de Astrof\'isica de Canarias.
\end{acknowledgements}

\end{document}